\documentclass[twoside,leqno,twocolumn]{article}
\usepackage{amsmath,bm}
\usepackage{amsfonts}
\usepackage{latexsym}
\usepackage{array}
\usepackage{graphicx}
\usepackage{url}
\usepackage{verbatim}
\usepackage{multirow}
\usepackage{threeparttable}
\usepackage{fixltx2e}
\usepackage[center]{caption} 
\usepackage{lscape}
\usepackage{rotating}
\usepackage{bigdelim}

\begin{document}

\title{\Large Paradigm shifts. Part II. Reverse Transcriptase.\\Analysis of reference stability and word frequencies.}
\author{Johannes Stegmann\footnote{Member of the Ernst-Reuter-Gesellschaft der Freunde, F\"orderer und Ehemaligen  der Freien Universit\"at Berlin e.V., Berlin, Germany, johannes.stegmann@fu-berlin.de}}
\date{}
\maketitle
\begin{abstract} The {\em reverse transcription} paradigm shift in RNA tumor virus research marked by the discovery of the {\em reverse transcriptase} in 1970 was traced using co-citation and title word frequency analysis. It is shown that this event is associated with a break in citation patterns and the occurrence of previously unknown technical terms.
\end{abstract}

\section{Introduction} \renewcommand*{\thefootnote}{\fnsymbol{footnote}}
\indent 
\indent In his annotations to the history of retrovirology in the standard textbook  {\em "Retroviruses"}  Peter K. Vogt links Thomas Kuhn's ideas of patterns of scientific revolutions (Kuhn, 1962) to several paradigm changes which happened during the development of retrovirology (Vogt, 1997, p. 16). One of the major and most prominent "revolution" was the detection of "reverse transcription"\footnote[2]{DNA synthesis from an RNA template}, more exactly, the identification of an RNA-dependent DNA polymerase in RNA tumor viruses (Temin and Mizutani, 1970; Baltimore, 1970). The idea of the occurrence of a DNA "provirus" intermediate in the infectious cycle of Rous sarcoma virus had been formulated already several years before (Temin, 1964). Because this hypothesis required the repeal of the so-called "central dogma" of molecular biology (one-way transfer of genetic information from DNA to RNA to protein) it was not taken seriously by the majority of the researchers in the discipline. The evidence of reverse transfers of genetic information (from RNA to DNA) was therefore indeed a scientific revolution. \\
\indent It is certainly of interest to reproduce unequivocal paradigm changes in science by bibliometric tools. Established techniques could be helpful in detecting ongoing transitions in scientific (sub-) fields. Kuhn himself suggested to study "shift(s) in the distribution of the technical literature cited in the footnotes to research reports" (Kuhn, 1962, p. ix) as possible indicators of paradigmatic changes in science. Following Kuhn's advice, Henry Small described in 1977 a paradigm shift in collagen research of the early 1970s (Small, 1977). The shift was brought about by the discovery of the collagen precursor molecule {\em procollagen} which had profound influences on basic biology, biochemistry and medicine (see Small, 1977, p. 147). Small traced the (pro-) collagen paradigm shift bibliometrically, applying co-citation analysis to the "footnotes", i.e. the cited references in the collagen literature set (Small, 1977). He did so by clustering all records of the appropriate publication years contained in the {\em Science Citation Index} year by year using definite citation/co-citation thresholds, followed by identifying the clusters with collagen literature and measuring the persistence of the cited references (called "reference stability") in succeeding years and three year intervals (Small, 1977). Analysing the latter Small found a citation "gap" in the three year interval 1971/1973, i.e. none of the cited references found in the 1971 literature set was also found in the 1973 set. This finding conforms to the detection of procollagen in 1971 indicating a different citation behavior in years following the procollagen discovery (Small, 1977).\\
\indent I could repeat and confirm Small's work processing collagen literature sets retrieved by appropriate keyword searches  (Stegmann, 2014). I used Small's technique of (co-) citation analysis and validated the results by exploring the frequencies of title words (Stegmann, 2014). \\
\indent The present study analysed the literature on {\em retroviruses} published 1966 to 1975 in order to find bibliometric reflections of the {\em reverse transcription} paradigm change which happened in 1970. The same methods as in Stegmann (2014) were used. First, a PUBMED search was conducted for the literature on retroviruses published 1966 to 1975 using appropriate MESH\footnote[3]{MESH = Medical Subject Headings is MEDLINE's hierarchichal thesaurus. MEDLINE is a subset of PUBMED. When indexed with MESH terms PUBMED records become MEDLINE records.} terms. A search in the publicly available database PUBMED is indispensable for retrieval of biomedical literature. Because PUBMED records lack cited references it is desirable to retrieve PUBMED equivalents with a cited references (CR) field. The Web of Science (WoS) database records contain such a CR field. WoS offers also an implementation of MEDLINE (WoS-M) which is equivalent to PUBMED's MEDLINE subset. So, the second step involved a search in WoS-M using a search profile equivalent to that used in the PUBMED search. However, records found in WoS-M lack the CR field and also WoS-specific record numbers (field UT in WoS records). To overcome this difficulty, the third step implied the application of the routine developed by Rotolo and Leydesdorff (2014). The routine identified WoS equivalents (if present) of WoS-M records which were downloaded in the fourth step. In further steps the downloaded WoS records were subjected to citation/co-citation and word frequency analyses. The downloaded PUBMED literature set (comprising about 25\% more records than its WoS pendant) was subjected to word frequency analyses.\\
\indent I will show (i) that the data of the reference stability analyis of the keyword-derived literature on {\em retroviruses} published 1966 to 1975 clearly indicate a profound change in citation behavior in the early 1970s which can be interpreted as paradigm shift in retrovirus research, and (ii) that text analysis confirms this view due to the occurrence of the phrase {\em "reverse transcriptase"}\footnote[4]{common term for "RNA-dependent DNA polymerase"}. \\
\indent I will use throughout the present paper the term {\em "retrovirus(es)"} instead of {\em "RNA tumor viruses"}. The latter was used prior to the discovery of reverse transcription as an inherent feature of RNA tumor viruses and is now used as an entry term (synonym) of the MESH term {\em "retroviridae"} which was introduced 1981 and comprises the oncogenic retroviruses and two minor groups of non-oncogenic retroviruses\footnote[5]{http://www.ncbi.nlm.nih.gov/mesh/68012190}.  

\section{Methods} \renewcommand*{\thefootnote}{\fnsymbol{footnote}}
\subsection{Online Retrieval and Download} \indent
\indent The search for the literature on {\em retroviuses} of the publication years 1966 - 1975 was performed on December 1, 2014, in the MEDLINE part of the Web of Science\footnote[6]{www.webofknowledge.com} (WoS-M). Because the WoS-M records do neither contain a CR (cited references) nor an UT (WoS identifier) field (see Introduction) the routine developed by Rotolo and Leydesdorff (2014) was applied. The Rotolo-Leydesdorff script (link given in the appendix of Rotolo and Leydesdorff, 2014) collects the UT field content, i.e. the unique number of document (hidden in the html code of each WoS-M record, provided it has a WoS counterpart) "on the fly" and writes it into a file. Than, this file containing the UTs of the WoS equilvalents of WoS-M records can be used to retrieve those WoS equivalents (Rotolo and Leydesdorff, 2014). By use of the Rotolo-Leydesdorff script I retrieved the WoS equivalents of the WoS-M retrovirus papers. These WoS counterparts were downloaded and subjected to analysis. \\
\indent The Rotolo-Leydesdorff routine is an {\em R-script}, i.e. one must have installed the software package R (R Core Team, 2013). \\
\indent The retrovirus-specific papers of the freely available database PUBMED\footnote[7]{www.ncbi.nlm.nih.gov/pubmed/} (more exactly its MEDLINE subset) was also retrieved and downloaded (July 21, 2014).

\subsection{Reference Stability Index} \indent
\indent The reference stability index (RSI) was calculated exactly as described by Small (Small, 1977; see also Stegmann, 2014). Applying certain citation/co-citation thresholds the cited core references of each year were determined, followed by determination of core references common to the compared publication years and the calculation of RSI\footnote[8]{RSI is the quotient of the number of references shared by the two compared years and the number of unique references of both years.}. Two year and three year intervals were analysed.

\subsection{Text Analysis} \indent
\indent Document frequencies of title words - i.e. the number of records with at least one instance of the word in their titles - were determined for each publication year. Stop words\footnote[9]{Stop word list for English texts, downloaded on April 19, 2007 from ftp://ftp.cs.cornell.edu/pub/smart/english.stop.} were excluded from the analysis. Because the WoS set and the PUBMED/MEDLINE set of retrovirus-specific papers are of different size, text analysis was performed on both sets. Frequency thresholds were applied, i.e. words had to occur in a certain number of papers. Comparisons were done for two year and three year intervals. The occurrences of "new" words (not present in the former year) were checked. Co-word analysis (Callon et al., 1986) was also performed. For each possible pair of a year's distinct title words its cosine similarity was calculated. Only term pairs with a cosine value equal to or greater than 0.25 (corresponding to an angle of about 75$^\circ$) were included in the subsequent frequency analysis. As with single words the occurrences of term pairs not present in the prior year of the two compared publication years were checked.

\subsection{Programming} \indent
\indent Extraction of record field contents, data analysis and visualisation were done using homemade programs and scripts for perl (version 5.14.2) and the software package R version 2.14.1 (R Core Team, 2013). All operations were performed on a commercial PC run under Ubuntu version 12.04 LTS.

\begin{table*}[htpb]\small
\caption{Retrovirus research 1966 - 1975}
\centering
\begin{tabular}{cccc}
\noalign{\smallskip}
\hline
\noalign{\smallskip}
Publication years & \multicolumn{2}{c}{No. of papers} & No. of distinct \\ 
                  &  WoS  & PUBMED   & cited references (WoS only) \\
\noalign{\smallskip} 
\hline
\noalign{\smallskip}
1966 & 242 & 352 & 2991 \\
1967 & 259 & 364 & 3867 \\
1968 & 353 & 481 & 4908 \\
1969 & 394 & 493 & 5003  \\
1970 & 448 & 654 & 5981  \\
1971 & 519 & 682  & 7282  \\
1972 & 677 & 847  & 8294  \\
1973 & 791 & 1045 & 9495  \\
1974 & 934 & 1115 & 11971  \\
1975 & 814 & 1127  & 11349 \\
1969 - 1975 & 5431 & 7160 & 44350 \\
\noalign{\smallskip}
\hline
\end{tabular}
\end{table*}


\section{Results and Discussion} \renewcommand*{\thefootnote}{\fnsymbol{footnote}}
\subsection{Retrieval} \indent
\indent Using the search string {\em MH:exp\footnote[1]{turns the "exploding" feature on, i.e. in one search step the literature on a general term together with the literature on all more specific terms hierarchically positioned ”under” the more general term is retrieved.}=Retroviridae OR MH:exp=Retroviridae Infections OR MH:exp=Retroviridae Proteins} 7160 records were found in the MEDLINE part of WoS (WoS-M) for the publication years 1966 - 1975. Of these records 5431 (75.9\%) were identified by the Rotolo-Leydesdorff routine to have a WoS counterpart. These 5431 WoS equivalents were downloaded and subjected to reference stability analysis and title (co-) word frequency determination. The search in PUBMED using the phrase {\em "Retroviridae"[Mesh]\footnote[2]{the "exploding" feature is automatically invoked.} OR "Retroviridae Infections"[Mesh] OR "Retroviridae Proteins"[Mesh]} limited to publication years 1966 to 1975 retrieved also 7160 records which were downloaded and subjected to title (co-) word frequency analysis. \\
\indent The distribution of these papers to their publication years is shown in Table 1. Also shown are the number of distinct references cited by the (WoS) papers of each year. The data presented in Table 1 show that the number of both, papers and references, have more than tripled from 1966 to 1974 or 1975. This is clearly more than the (paper) growth of the whole WoS and PUBMED databases which increase from 1966 to 1974/1975 by a factor of about 1.5 to 1.7 and 1.3, respectively (data not shown), indicating that retrovirus research experienced a major boost in the early 1970s. 

\subsection{Reference Stability} \indent
\indent Highly cited references can be regarded as symbols of relevant scientific concepts, and co-citation of those references can inform about the strenghts of links between the concepts (Small, 1977). Thus, to be able to detect paradigmatic changes of concepts, it is reasonable to apply certain citation/co-citation thresholds in reference stabiliy analysis (Small, 1977). The results of the reference stability analysis under different thresholds are shown in Table 2 and Table 3 (analysing two year intervals) and Table 4 (analysing three year intervals). Table 2 and Table 3 list for each year the number of core references and the number of core references common to two successive years as well as the resulting RSI value. The 15/11 threshold is obviously too high because no core references were found for the publication year 1967 (see Table 2, first row). Certainly, no "trend" can be read from the data displayed in Table 2 and Table 3. There is not a definite decrease in RSI to the years of change in retrovirus research (1970, 1971). Thus, a major shift in retrovirus concepts can not be inferred from Table 2 and Table 3.

\begin{table*}[htpb]\small
\caption{Reference stability of core references in Retrovirus research 1966 - 1970: two year intervals.}
\centering
\begin{threeparttable}

\begin{tabular}{cccccccccl}
\hline
\noalign{\smallskip}
Citation/Co-citation & \multicolumn{9}{c}{Number of core and shared core references and reference stability indexes of two year intervals} \\
 thresholds & \\ 
 & 1966 & sh\tnote{1}/RSI\tnote{2} & 1967 & sh/RSI & 1968 & sh/RSI & 1969 & sh/RSI & 1970 \\
\noalign{\smallskip}
\hline
\noalign{\smallskip}
15/11 & 3 & -/- & 0 & -/- & 4 & 2/0.20 & 8 & 6/0.33 & 16 \\
15/8 & 8 & 2/0.22 & 3 & 2/0.25 & 7 & 6/0.35 & 16 & 9/0.28 & 25 \\
11/9 & 7 & 2/0.18 & 6 & 5/0.42 & 11 & 3/0.12 & 18 & 8/0.21 & 29  \\
10/8 & 11 & 4/0.27 & 8 & 5/0.26 & 16 & 8/0.25 & 24 & 13/0.23 & 46  \\
\noalign{\smallskip}
\hline
\end{tabular}
\begin{tablenotes}
\footnotesize
\item[1] sh: number of core references shared by both years.
\item[2] RSI: reference stability index (see Methods).
\end{tablenotes}
\end{threeparttable}
\end{table*}

\begin{table*}[htpb]\footnotesize
\caption{Reference stability of core references in Retrovirus research 1970 - 1975: two year intervals.}
\centering
\begin{threeparttable}
\begin{tabular}{cccccccccccc}
\hline
\noalign{\smallskip}
Citation/Co-citation  & \multicolumn{11}{c}{Number of core and shared core references and reference stability indexes of two year intervals} \\
        thresholds & \\
 & 1970 & sh\tnote{1}/RSI\tnote{2} & 1971 & sh/RSI & 1972 & sh/RSI & 1973 & sh/RSI & 1974 & sh/RSI & 1975 \\
\noalign{\smallskip}
\hline
\noalign{\smallskip}
15/11 & 16 & 10/0.26 & 32 & 13/0.26 & 31 & 16/0.21 & 60 & 25/0.22 & 77 & 25/0.27 & 40 \\
15/8 & 25 & 18/0.35 & 44 & 15/0.25 & 31 & 21/0.21 & 89 & 55/0.34 & 127 & 43/0.30 & 61 \\
11/9 & 29 & 17/0.24 & 60 & 31/0.13 & 72 & 41/0.31 & 103 & 60/0.32 & 145 & 42/0.24 & 70 \\
10/8 & 46 & 26/0.25 & 85 & 39/0.31 & 80 & 47/0.27 & 138 & 45/0.37 & 200 & 71/0.30 & 108 \\
\noalign{\smallskip}
\hline
\end{tabular}
\begin{tablenotes}
\footnotesize
\item[1] sh: number of core references shared by both years.
\item[2] RSI: reference stability index (see Methods).
\end{tablenotes}
\end{threeparttable}
\end{table*}

\begin{table*}[htpb]\small
\caption{Reference stability of core references in retrovirus research 1966 - 1975: three year intervals.}
\centering
\begin{threeparttable}
\begin{tabular}{ccccccccl}
\hline
\noalign{\smallskip}
Citation/Co-citation  &  \multicolumn{8}{c}{Number of shared core references and RSIs of three year intervals\tnote{1,2}}\\
 thresholds &    \\  
 & 1966/1968 & 1967/1969 & 1968/1970 & 1969/1971 & 1970/1972 & 1971/1973 & 1972/1974 & 1973/1975 \\
\noalign{\smallskip}
\hline
\noalign{\smallskip}
15/11  & 0/0.00 & 0/0.00 & 2/0.11 & 5/0.14 & 2/0.04 & 10/0.12 & 17/0.19 & 13/0.15 \\
15/8  & 3/0.25 & 2/0.12 & 6/0.23 &  10/0.20 & 2/0.04 & 17/0.15 & 23/0.17 & 27/0.22 \\
11/9  & 3/0.20 & 3/0.14 & 4/0.11 &  8/0.11 & 6/0.06 & 19/0.13 & 39/0.22 & 25/0.17 \\
10/8  & 3/0.12 & 5/0.19 & 9/0.17 &  15/0.16 & 8/0.07 & 28/0.14 & 48/0.21 & 45/0.22 \\
\noalign{\smallskip}
\hline
\end{tabular}
\begin{tablenotes}
\footnotesize
\item[1] Values are denoted in the sequence sh/RSI (see Table 2 and Table 3).
\item[2] For numbers of year-specific core references see Table 2 and Table 3.
\end{tablenotes}
\end{threeparttable}
\end{table*}

\begin{table*}[htpb]\small
\caption{Core references in retrovirus research shared by papers published 1970 and 1972.}
\centering
\begin{tabular}{cccl}
\hline
\noalign{\smallskip}
\multicolumn{3}{c}{Citation/Co-citation thresholds}  & Cited references\ \\
\noalign{\smallskip}
\hline
\noalign{\smallskip}
\multirow{8}{*}{10/8} \rdelim\}{8}{1pt} & \multirow{6}{*}{11/9} \rdelim\}{6}{1pt} & \multirow{2}{*}{15/11 and 15/8} \rdelim\}{2}{1pt} &  BALTIMORE D, 1970, NATURE, V226, P1209 \\
    &  & & TEMIN HM, 1970, NATURE, V226, P1211 \\
          \cline{4-4} 
\noalign{\smallskip}
  &   &     &     HUEBNER RJ, 1966, P NATL ACAD SCI USA, V56, P1164 \\ 
  &   &     &     HARTLEY JW, 1965, P NATL ACAD SCI USA, V53, P931 \\
  &   &     &     HARTLEY JW, 1966, P NATL ACAD SCI USA, V55, P780\\
  &   &     &     HARTLEY JW, 1969, J VIROL, V3, P126 \\
     \cline{4-4} 
\noalign{\smallskip}
    &   & &  DUFF RG, 1969, VIROLOGY, V39, P18 \\
  &   &  &     SPIEGELM.S, 1970, NATURE, V227, P563 \\

\noalign{\smallskip}
\hline
\end{tabular}
\end{table*}

\indent In contrast to Table 2 and Table 3 , Table 4 (displaying numbers of shared core references and resulting RSIs for three year intervals) clearly shows a citation "groove" in the interval 1970/1972. Here, each of the four RSI series has its minimal value which is close to zero (between 0.04 and 0.07) (see Table 4, column 6)\footnote[3]{The "0"-values of the intervals 1966/1968 and 1967/1969 (Table 2, row 1) are due to the very restrictive constraints of the 15/11 threshold; see also above.}.
Obviously, the reference stability in retrovirus research experienced a major break in the interval 1970/1972. According to Small's interpretation of the citation gap found in collagen research as a "conceptual shift" (Small, 1977) we may conclude from the data in Table 4 that - beginning in 1970 - (a) new concept(s) is/are introduced into retrovirology replacing former basic views of the specialty.

\indent That the minimal RSI values are close to but not equal to zero may depend on (i) the citation/co-citation thresholds applied, (ii) the time intervals investigated, and/or (iii) the publication date of the paper(s) inducing the paradigm shift. Papers and cited references are analysed on the basis of entire publication years. So, a groundbreaking paper published early in the year in a journal which is quickly absorbed by the scientific community might already be frequently cited in the same year. We see that under the thresholds 15/11 and 15/8 two cited references persist from 1970 to 1972. Application of the 11/9 threshold results in additional 4 shared core references, and under the 10/8 threshold two more core references are shared by the literature sets of 1970 and 1973 (see Table 4). Table 5 displays the cited reference strings of these core references. The main two cited references represent the famous {\em Nature} papers of Temin and Mizutani (1970) and Baltimore (1970), persisting also under the strongest threshold used in this investigation. They clearly stand for the new paradigm {\em reverse transcription} which entered 1970 not only (retro-) virology but also the whole field of biomedicine. The two papers were published side by side in June 1970. Still in the same year both papers belong to the group of most often co-cited papers. In each of the subsequent years (up to 1975) they are the top co-cited papers and are also mostly the top cited papers (data not shown). The Temin/Mizutani paper and the Baltimore paper of 1970 are new at that time and they are found within the core references of the 1970 paper set. Therefore, it is certainly advisable for an investigation looking for possible shifts in scientific concepts also to check whether new (highly) cited references appear. The RSI values of the two year interval 1970/1971 are similar to those of the other two year intervals (see Table 2 and Table 3) because other (older; not shown) references are present in both years; thus they do not point to a possible paradigm change. Within the three year interval 1970/1972 the old references are more loosely connected and appear only when thresholds are lowered (see Table 5). In the subsequent years they do not have their previous significance and some of them even vanish from the list of core references (not shown). \\
\indent In summary we can state that the reference stability analysis developed by Henry Small is able to detect the radical change in retrovirology caused by the discovery and the proof of reverse transcription. Expert knowledge doesn't seem to be  obligatory provided a comprehensive subject-specific literature search is the basis of the study. Analysis of three year intervals is certainly more adequate than shorter periods. The RSI can point to changes in citation behavior indicating profound shifts in the specialty. If so, then of course more data must be collected including experts' opinions.



\subsection{Text Analysis} \indent

\begin{table*}[htpb]\small
\caption{New title words in retrovirus research: comparing two year and three year intervals 1969 - 1972.}
\centering
\begin{threeparttable}
\begin{tabular}{cclc|clc}                            
\hline
\noalign{\smallskip}
\multirow{3}{*}{Former publication year}  & \multicolumn{6}{l}{\hspace*{4cm}New title words in later publication year\tnote{1,2,3,4}} \\
                        & \multicolumn{3}{c}{1971} & \multicolumn{3}{c}{1972}  \\
                        & PUBMED & & WoS & PUBMED & & WoS \\
\noalign{\smallskip}
\hline
\noalign{\smallskip}
\multirow{17}{*}{1969}  & 1.9 & DEOXYRIBONUCLEIC & - & & & \\
                        & 1.7 & GENOME & 2.1 & & & \\
                        & 1.3 & POLYMERASES  & 1.5  & & & \\
                        & 1.2 & INHIBITORS   & 1.3  & & & \\
                        & -   & SPECIFICITY  & 1.3  & & & \\
                        & -   & CHRONIC      & 1.2  & & & \\
                        & -   & COMPOSITION  & 1.2  & & & \\
                        & -   & DERIVED      & 1.2  & & & \\
                        & 1.2 & \textbf{REVERSE}      & 1.0  & & & \\
                        & 1.0 & ENZYME       & 1.2  & & & \\
                        & 1.0 & \textbf{TRANSCRIPTASE} & 1.0  & & & \\
                        & -   & PRODUCT      & 1.2  & & & \\
                        & -   & TURKEYS      & 1.2  & & & \\
                        & -   & FUSION       & 1.0  & & & \\
                        & -   & METHYLCHOLANTHRENE & 1.0  & & & \\
                        & -   & NUCLEOSIDE   & 1.0  & & & \\
                        & -   & REQUIREMENT  & 1.0  & & & \\
         \cline{2-7} 
\noalign{\smallskip}

\multirow{9}{*}{1970}   & 1.2 & INHIBITORS   & 1.3  & 1.8 & \textbf{REVERSE} & 2.1 \\
                        & - & COMPOSITION    & 1.2  & 1.2 & \textbf{TRANSCRIPTASE} & 1.6 \\
                        & - & RETICULOENDOTHELIOSIS & 1.2 & 1.7 & BREAST & 1.5 \\
                        & 1.2 & \textbf{REVERSE}   & 1.0     & 1.3 & DERIVATIVES  & 1.5 \\
                        & 1.0 & \textbf{TRANSCRIPTASE}    & 1.0  & 1.2 & ONCORNAVIRUSES & 1.0 \\
                        & -   & DISEASES   & 1.0 & 1.0 & RIFAMYCIN & 1.2 \\
                        & -   & INCREASED     & 1.0  & - & SEQUENCES & 1.0 \\
                        & -   & NATURE   & 1.0  & 1.0 & TEMPERATURE  & -\\
                        & - & RIFAMPICIN   & 1.0  & & & \\  
\noalign{\smallskip}
\noalign{\smallskip}
\hline
\end{tabular}
\begin{tablenotes}
\footnotesize
\item[1] Only words which appeared in at least 1\% of the later year's papers were included.
\item[2] Words are sorted top down according to frequency if possible.
\item[2] The columns headed "PUBMED" and "WoS" give the word frequency in percent of papers in the database in the respective year.
\item[4] The comparison of 1969 and 1970 is not displayed. Here, under the threshold used only three words were new in both databases: BITTNER, GUINEA, LIGHT.
\end{tablenotes}
\end{threeparttable}
\end{table*}

\begin{table*}[htpb]\tiny
\caption{New title co-words in retrovirus research: comparing two year and three year intervals 1969 - 1972.}
\centering
\begin{threeparttable}
\begin{tabular}{ccl|clc}                            
\hline
\noalign{\smallskip}
\multirow{3}{*}{Former publication year}  & \multicolumn{5}{l}{\hspace*{4cm}New title co-words in later publication year\tnote{1,2,3,4}} \\
                         & \multicolumn{2}{c}{1971} & \multicolumn{3}{c}{1972}  \\
                         & PUBMED &                &  PUBMED &  & WoS \\
\noalign{\smallskip}
\hline
\noalign{\smallskip}
\multirow{17}{*}{1969}  & 6.8 & MICE/VIRUS &  & & \\
                        & 3.1 & CELLS/INFECTED & & & \\
                        & 2.9 & MAMMARY/MOUSE  & & & \\
                        & 2.8 & FELINE/LEUKEMIA  &  & & \\
                        & 2.8  & MOUSE/TUMOR  &  & & \\
                        & 1.9  & ACID/DEOXYRIBONUCLEIC   &   & & \\
                        & 1.9  & CHICKENS/DISEASE  &   & & \\
                        & 1.9  & INFECTED/MICE      &  & & \\
                        & 1.7  & ANTIGENS/TUMOR   &   & & \\
                        & 1.6  & AVIAN/STRAIN      &  & &  \\
                        & 1.6  & FRIEND/MICE &   & &  \\
                        & 1.3  & FRIEND/INFECTED  &  & & \\
                        & 1.2  & CELL/LINES      &  & &  \\
                        & 1.0  & ACID/POLYMERASE &   & &  \\
                        & 1.0  & CHARACTERIzATION/ISOLOATION  & & & \\
                        & 1.0  & DEOXYRIBONUCLEIC/POLYMERASE    & & & \\
                        & 1.0  & \textbf{REVERSE/TRANSCRIPTASE}  &  & & \\
         \cline{2-6} 
\noalign{\smallskip}

\multirow{16}{*}{1970}  & 6.0 & MOUSE/VIRUS   & - & AVIAN/VIRUS & 6.1  \\
                        & 2.3 & FRIEND/LEUKEMIA & - & TUMOR/VIRUSES  & 5.9 \\
                        & 1.9 & INFECTED/MICE & 5.2 & MOUSE/VIRUSE & 5.6 \\
                        & 1.7 & ANTIGENS/TUMOR    & - & CELLS/MURINE  & 4.9 \\
                        & 1.6 & ANTIGENS/SPECIFIC  & 3.9 & CELLS/ROUS & 4.0  \\
                        & 1.6  & AVIAN/STRAIN    & 3.7 & RAUSCHER/VIRUS  & - \\
                        & 1.6   & FRIEND/MICE      & 3.5 & SARCOMA/TRANSFORMED & 3.7 \\
                        & 1.6   & PARTICLES/TYPE   & 2.9 & LEUKEMIA/MOUSE  & - \\
                        & 1.3  & FRIEND/INFECTED   & - & ACID/RIBONUCLEIC & 2.2 \\  
                        & 1.2  & CELL/LINES  & 2.0 & ROUS/TRANSFORMED & - \\  
                        & 1.0  & ACID/POLYMERASE    & 1.9 & CELL/LINE & - \\  
                        & 1.0  & CHARACTERIZATION/ISOLATION & 1.6 & ACID/POLYMERASE & 1.9 \\  
                        & 1.0  & \textbf{REVERSE/TRANSCRIPTASE}  & 1.4  & CELL/SURFACE & 1.8 \\  
                        &      &                        &  -   & DEOXYRIBONUCLEIC/POLYMERASE & 1.8 \\
                        &      &                        & 1.7  & CELL/CULTURES & - \\
                        &      &                        & 1.2  & \textbf{REVERSE/TRANSCRIPTASE} & 1.6 \\

\noalign{\smallskip}
\noalign{\smallskip}
\hline
\end{tabular}
\begin{tablenotes}
\footnotesize
\item[1] Listed are only those co-word sets which contain the term pair {\em reverse/transcriptase}.
\item[2] Only co-words which appeared in at least 1\% of the later year's papers were included.
\item[3] Co-words are sorted top down according to frequency if possible.
\item[4] The columns headed "PUBMED" and "WoS" give the co-word frequencies in percent of papers in the respective database in the respective year.
\end{tablenotes}
\end{threeparttable}
\end{table*}

\indent Only titles were included in the text analysis, because none of the WoS records of 1969 - 1975 dealing with retrovirology contains an abstract, and the proportion of abstracts contained in the corresponding PUBMED records is with 4.6\% to 18.9\% for the years 1966 to 1974 rather small. Only in 1975 more than half of the PUBMED papers (63.1\%) contain abstracts (not shown). 

\indent Table 6 and Table 7 present some data of the title word frequency analyses from both databases, PUBMED and WoS. Because the results of the reference stability analysis (see preceding subsection) suggest the beginning 1970s as literal "turn-around" in retrovirology (discovery of reversed transcription) only data from 1969 and later are shown.
 
\indent Table 6 shows new occurrences of single title words, and Table 7 shows new occurrences of co-words, i.e. word pairs occurring in titles. In all displayed sets the (bold) terms {\em "reverse"} and {\em "transcriptase"} occur, either as single words or as strongly connected word pair, i.e. {\em reverse} and {\em transcriptase} are new words in 1971 (as compared with the word sets of 1969 and 1970) and in 1972 (as compared with 1970). Both terms ("reverse" and "transcriptase" and similar words as "transcript" and "transcription") are completely absent from the 1970 title word set in both databases even if no frequency threshold is applied (not shown, but see Figure 1) This demonstrates on the one hand the duty to apply - in an unbiased investigation (not knowing the results a priori) - more in-depth text analyses, including identification of multi-word phrases and obtaining expert's testimony, if one hopes to detect something which could emerge as a paradigm shift. For example, the two key papers of Temin/Mizutani and Baltimore heralding the paradigm change in retrovirus research already appeared in the middle of the publication year 1970 (see above). The two papers speak of "RNA-dependent DNA polymerase". To identify this phrase as new (in 1970) an analysis of phrases comprising four words would be necessary. On the other hand, we see that the technical phrase {\em reverse transcriptase} has relatively fast established as the label of the paradigm change in virology (see Table 6 and Table 7). This is confirmed by Figure 1 which displays the cumulated title frequencies of the phrases built of the word {\em reverse} followed by any words with the word stem {\em transcr}, e.g. {\em reverse transcriptase}, {\em reverse transcription} etc.

\indent In summary, we can say that even the simple form of text analysis used in this investigation found key phrases indicating the {\em reverse transcription} paradigm shift in retrovirology which started in the early 1970s.

\begin{figure}[htpb]
\centerline{
\includegraphics[height=8.0cm]{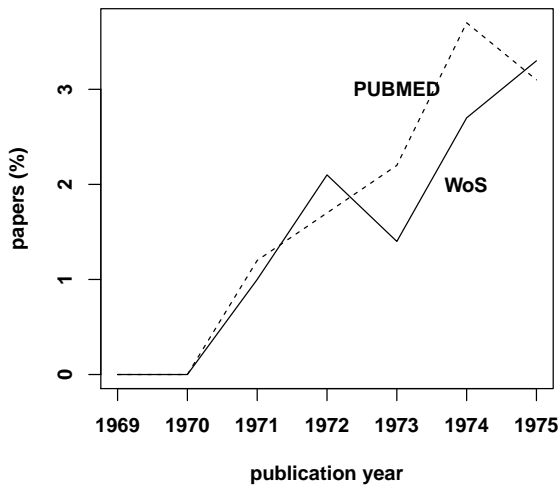}}
\caption{Title frequencies of the word phrase {\em reverse transcr\textsuperscript{*}} in retrovirus research papers 1969 - 1975.}
*\small includes {\em transcriptase, transcriptases, transcription, transcribed, transcript, transcripts}.
\end{figure}


\section{Conclusion} \indent
\indent The study presented here has shown that the paradigm shift in (retro-) virology caused by the discovery of the enzyme {\em reverse transcriptase} in 1970 can be traced applying Henry Small's method of reference stability analysis and methods of text analysis to literature sets retrieved by appropriate keyword searches.


\end{document}